\documentclass[%
 reprint,
 amsmath,amssymb,
 aps,
prl,
]{revtex4-1}

\usepackage{graphicx}
\usepackage{dcolumn}
\usepackage{bm}
\usepackage{color}
\usepackage{floatrow}
\usepackage[label font=bf,labelformat=simple]{subfig}
\usepackage{caption}
\begin{document}

\preprint{APS/123-QED}

\title{Recovering noise-free quantum observables}
  
\author{Matthew Otten}
\author{Stephen Gray}%
 \affiliation{%
 Center for Nanoscale Materials, Argonne National Laboratory, Lemont, Illinois, 60439
 }%

\date{\today}

\begin{abstract}
We introduce a technique for recovering noise-free observables
in
noisy quantum systems by combining the results of many
slightly different experiments.
Our approach is applicable to a variety of quantum systems
but we illustrate it with applications to quantum computing and
quantum sensing.
The approach corresponds to repeating
the same quantum evolution many times with known variations on the 
underlying systems' error properties, e.g. the spontaneous emission
and dephasing times, $T_1$ and $T_2$.
As opposed to standard quantum error correction methods, which have an
overhead in the number of qubits (many physical qubits must be added 
for each logical qubit)  our method has only an overhead in number of
evaluations, allowing the overhead to, in principle,  be hidden via 
parallelization. 
We show
that the effective spontaneous emission, $T_1$, and dephasing, $T_2$, times can be increased using this method
in both simulation and 
experiments 
on an actual quantum computer. 
We also show how to correct more
complicated entangled states and how Ramsey fringes relevant to
quantum sensing can be significantly extended in time.

\end{abstract}

\pacs{Valid PACS appear here}
\maketitle


Quantum information science is rapidly evolving 
due to advances in quantum computing, communication, and sensing.
Quantum computing, for example, has the 
possibility to show exponential
speedup in areas such as prime number factoring~\cite{shor1999polynomial} and
quantum chemistry~\cite{lanyon2010towards} and quantum sensing
has the potential to be able to be far more sensitive than classical
sensing~\cite{degen2017quantum}.
However, in all quantum information applications, decoherence (or information
loss) is a serious problem in 
advancing beyond elementary demonstrations.
Quantum error correction is a way to extend the lifetime of
quantum information by encoding a single logical qubit into
many physical qubits~\cite{nielsen2010quantum}. 
This introduces both space and time overheads 
owing to the additional number of physical qubits required and the
additional gate operations. Furthermore, it is not clear how standard
quantum error correction could be used in complicated quantum sensors.

\begin{figure}
  \centering
  \includegraphics[width=\columnwidth]{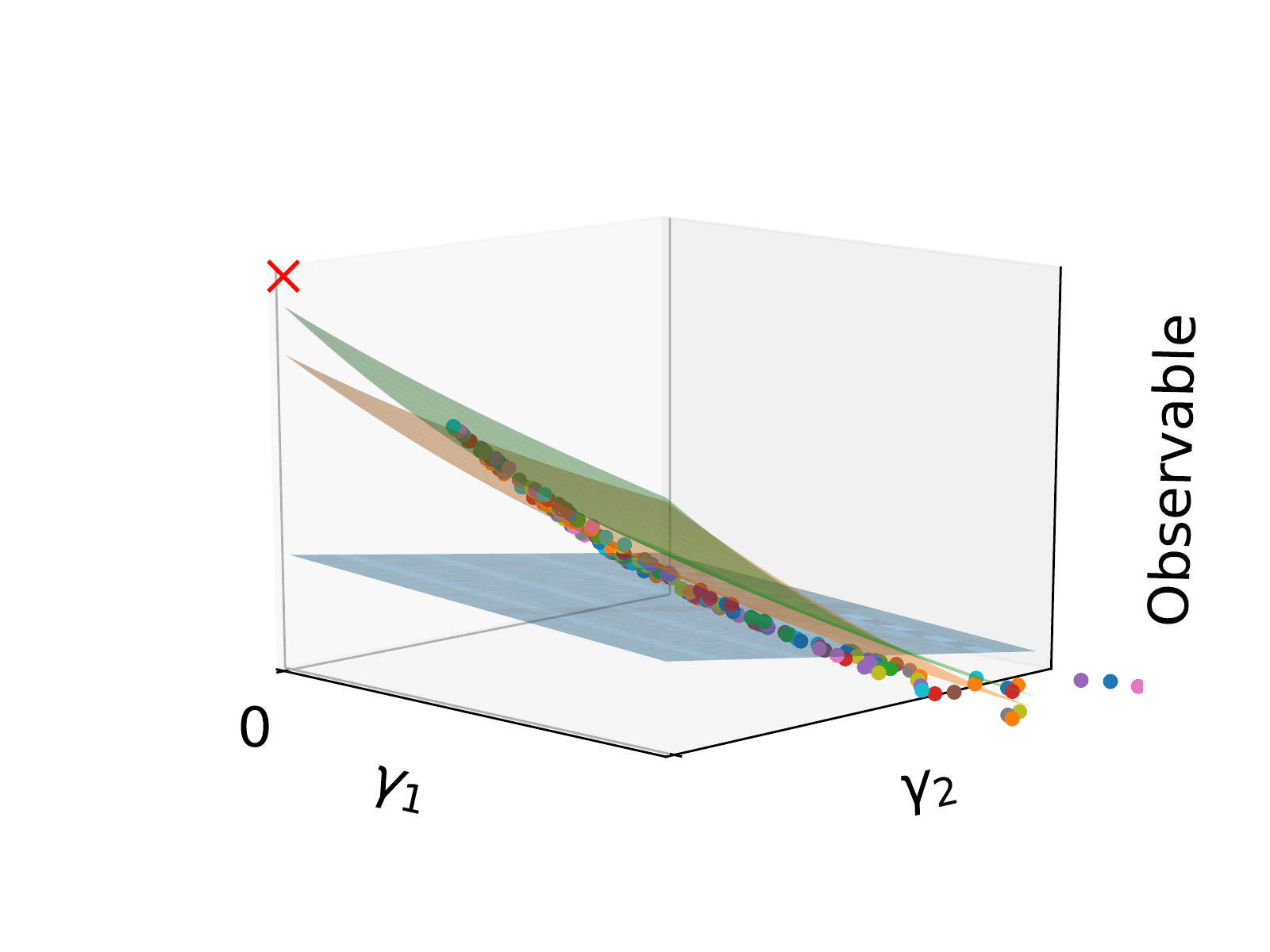}
  \caption{An example of a `hypersurface' fit to many experiments with slightly
    different noise parameters, $\gamma_1$ and $\gamma_2$. The red `X'
    represents the noise-free observable. The blue, orange, and green surfaces are first,
    third, and fourth order fits, respectively. Note that this plot is zoomed;
    many other observable measurements are outside of the visible region.}\label{hypersurface_fit}
\end{figure}

In this Letter, we describe a
technique for recovering observables from  a quantum evolution by
repeating the evolution with slightly different noise
characteristics and combining those results to obtain an estimate of the 
noise-free answer without the need of additional quantum hardware.  
Our approach bares some similarities with interesting recent work by Gambetta and 
coworkers~\cite{temme2017error,kandala2018extending} involving one noise
parameter, a simple model for observable variation with error, and 
Richardson extrapolation. The approach we develop may be viewed as a 
multi-dimensional
generalization not reliant on Richardson extrapolation. 
We should note that we have recently used such a simple error model
to develop a different error correction scheme that requires
the ability to signficantly reduce error on individual
qubits via, e.g., quantum error correction \cite{x-prl}. 
The approach presented here,
which does not rely on  quantum error correction, 
requires many, slightly different runs to be
completed; these can be done in time (repeating the evolution many times in
sequence) or in parallel (many separate quantum systems undergo the same
evolution at the same time).  
The ability to trade overhead between space
and time also has advantanges over standard quantum error correction. 
Our approach can be used in quantum algorithms,
quantum sensing, and even
general quantum experiments where the decoherence time is too short to obtain
high-quality signals. The only requirement is that the evolution can be
repeated with different, well-characterized noise sources.

Consider a quantum system with one or more subsystems (e.g., qubits), each
undergoing a small set of dominant noise processes
parameterized by a set of noise rates
$\{\gamma_i\}$. 
Suppose that this system repeatedly undergoes a given evolution
(such as a sequence of quantum gates or interaction with a magnetic field) with
varying values of the noise rates 
$\{\gamma_i\}$. 
These repetitions can be accomplished sequentially on one 
system (consuming time)
or in parallel on a set of  systems (consuming space) or in
some combination.
Combining all results, we construct a hypersurface 
embedded
in a space where one axis represents the results of a measurement 
and each of the
other axes represent the effect of a noise rate parameter. 
The form of
the hypersurface is obtained via the Taylor expansion of the
quantum system's evolution operator (see Appendix). 
This hypersurface yields an estimate of the noise-free observable and
gives general information about the effect of each noise rate.

To be concrete, consider a single qubit with only amplitude
damping noise. We repeatedly apply some evolution to this qubit,
but each time the noise rates are different. 
Let $\gamma^{[j]}_1$  be the amplitude
damping rate for the $j$th
repetition and $\langle A \rangle^{[j]}$ be the corresponding measured observable. 
Restricting our observable model to third order in $\gamma_1$, 
we would solve
(by standard least-squares procedures)
\begin{gather}\label{hs_lsq}
  \begin{bmatrix}
    1 & \gamma^{[1]}_1 & (\gamma^{[1]}_1)^2 & (\gamma^{[1]}_1)^3  \\
    1 & \gamma^{[2]}_1 & (\gamma^{[2]}_1)^2 & (\gamma^{[2]}_1)^3 \\
    1 & \gamma^{[3]}_1 & (\gamma^{[3]}_1)^2 & (\gamma^{[3]}_1)^3  \\
    \vdots & \vdots & \vdots & \vdots 
  \end{bmatrix}
  \begin{bmatrix}
      A_0  \\
      A_1 \\
      A_{11} \\
      A_{111}  
  \end{bmatrix}
  =
  \begin{bmatrix}
    \langle  A \rangle^{[1]}\\
    \langle  A \rangle^{[2]}\\
    \langle  A \rangle^{[3]}\\
    \vdots
  \end{bmatrix},
\end{gather}
to obtain $(A_0,A_1,A_{11},A_{111})$ 
and the observable as a function of the noise rate is given by
$\langle A \rangle \approx$ $A_0 + A_1 \gamma_1 + A_{11} \gamma_1^2
+ A_{111} \gamma_1^3$. In this case the hypersurface is just a cubic
curve with intercept $A_0$ being the desired noise-free
value.
The formalism easily extends to higher orders and to many noise
parameters, potentially from many qubits. 
For example, 
using a single qubit with a spontaneous emission rate, $\gamma_1$ and pure
dephasing rate, $\gamma_2$, the $j$th row of our matrix would 
be $[1$
$\gamma_1^{[j]}$ $\gamma_2^{[j]}$ $(\gamma_1^{[j]})^2$ $\gamma_1^{[j]}\gamma_2^{[j]}$
$(\gamma_2^{[j]})^2]$, where we truncate to second order. 
Figure~\ref{hypersurface_fit} displays hypersurfaces (now 2D surfaces) for 
such a system with two noise parameters. 
A red `X' is the noise-free
solution. 
As model order is 
increased, the surfaces  better fit the data and extrapolate
closer to the noise-free limit. For many qubits with many noise rates, the
result is a high-dimensional hypersurface.

\begin{figure}
  \centering
  \sidesubfloat[]{\includegraphics[width=\columnwidth]{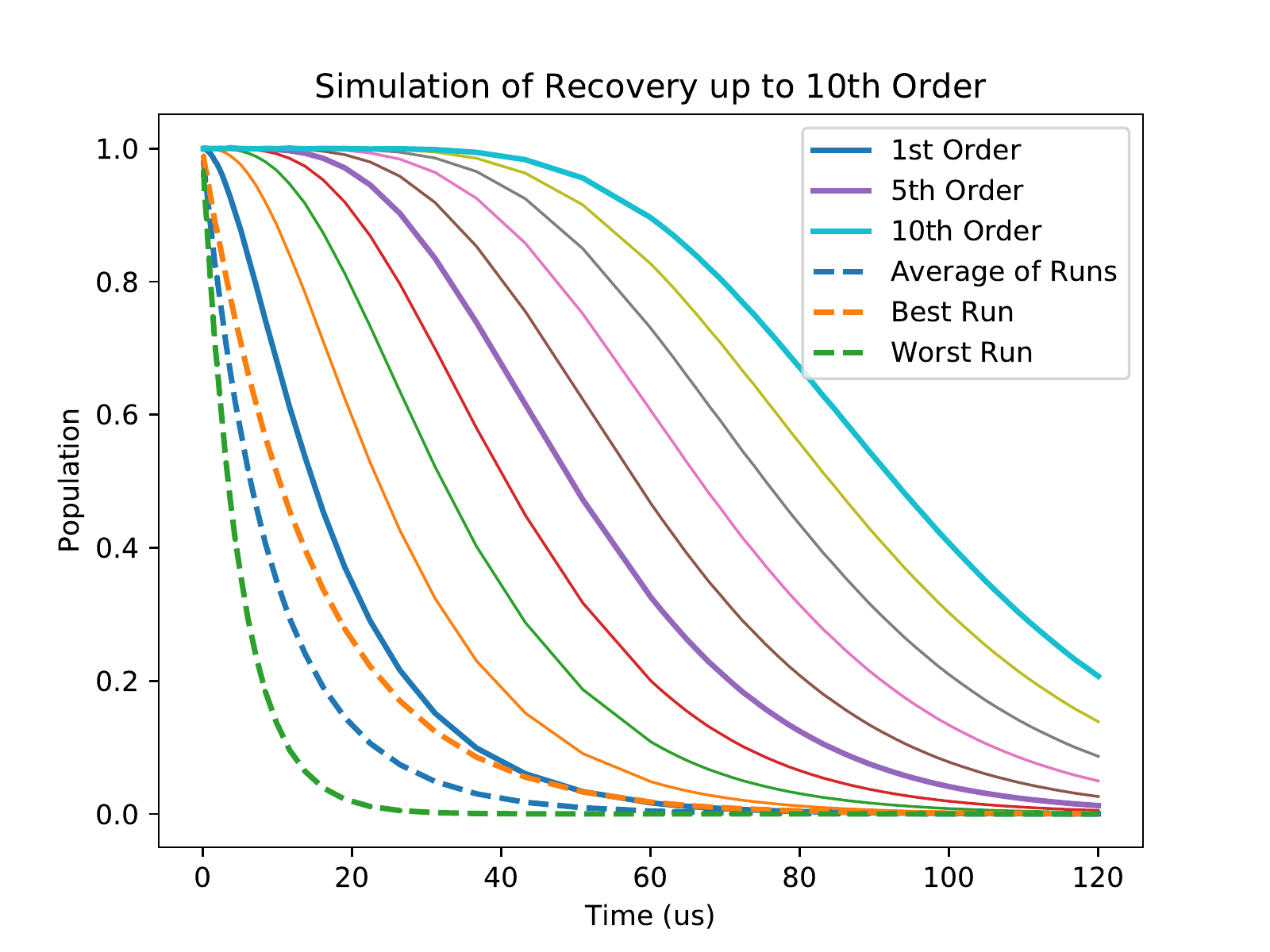}\label{t1_sim_10th}}
  \hfil
  \sidesubfloat[]{\includegraphics[width=\columnwidth]{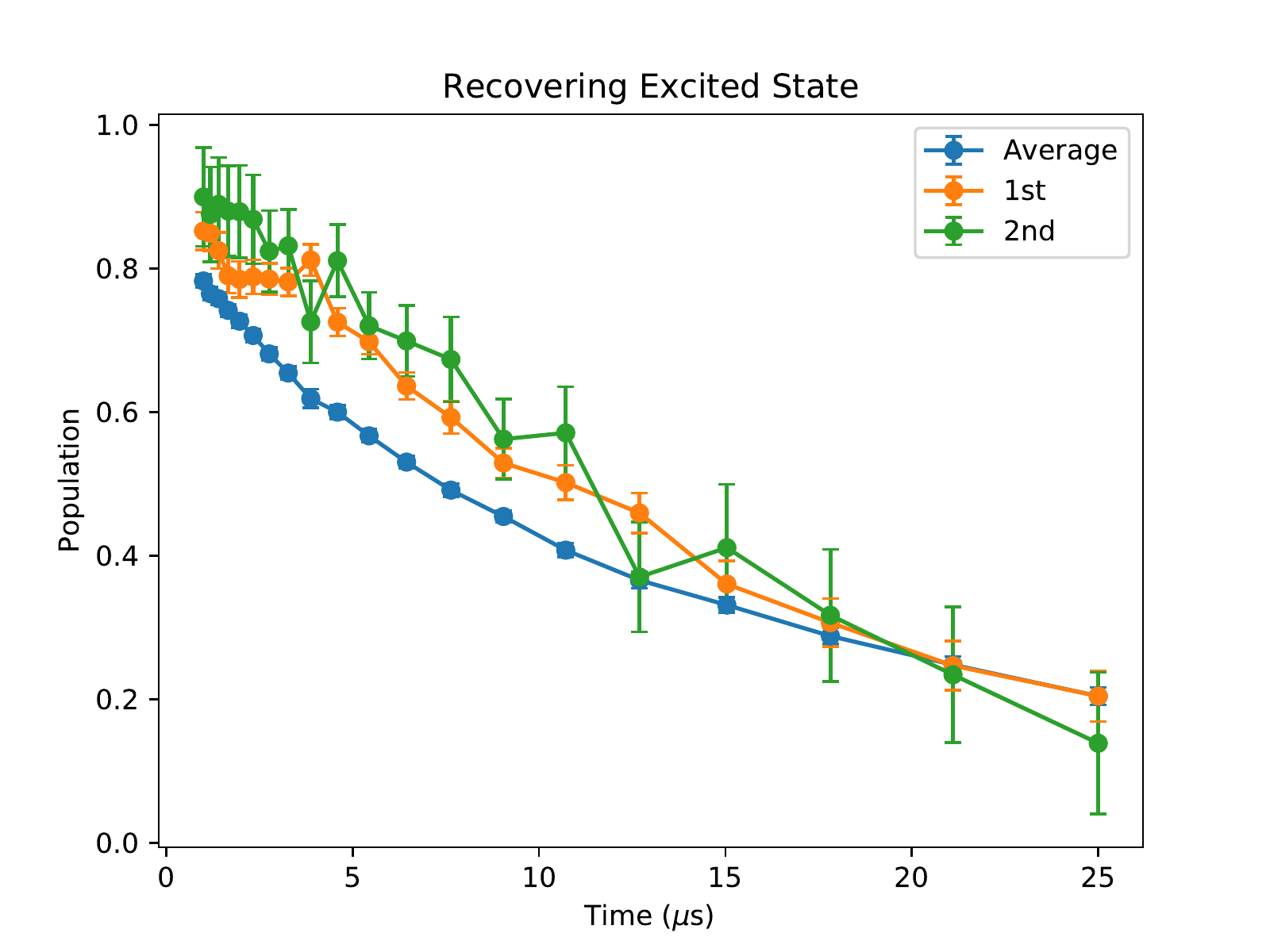}\label{t1_experiment}}
  \caption{Recovery of the population in a relaxation time experiment. (a)
    Simulated data from 450 simulations  with random $T_1$ times. (b)
    Experimental results on Rigetti's 8 qubit quantum computer, 
      Agave~\cite{reagor2018demonstration}. 45 separate repetitions are made at
      each time with experimentally determined $T_1$ times.}\label{t1_figs}
\end{figure}

We first demonstrate this method for a single qubit using a simple relaxation
time experiment. We
excite the qubit into the $| 1 \rangle$ state, wait some amount of time, and
then measure what state it is in; repeating this many times allows a probability
 of remaining in (or fractional population of) the excited state
to be obtained.
Without noise, each measurement result (and the fractional population of the
excited state) would be unity for
all times. Due 
to the amplitude damping noise, the population will decay to zero with
characteristic time $T_1$. We first show this in simulation, where we select
random $\gamma_1$ values uniformly in a range representing $T_1$ times
between 5 $\mu$s and 15
$\mu$s. All simulations are numerical solutions of the  
Lindblad master
equation~\cite{otten-prb-2015,otten-pra-2016} and 
utilize the high-performance open quantum systems solver QuaC~\cite{QuaC:17}.
The results are shown in Fig.~\ref{t1_sim_10th}, where we plot the
best, worst, and average
evolutions over 450 repetitions, and the recovered populations using
equation~\eqref{hs_lsq}, up to tenth order. 
The procedure is applied at
specific times, with knowledge only of the measured observables at that time. By
recovering at many different times $t$, we obtain the full evolution.
Every order, from first until tenth, is
better than the best run; increasing the order increases the quality of the
recovery. Take, for example, the value of the observable at 60
$\mu$s. At this
point, the average population is 0.0045; essentially all of the state has
decayed away. The first order recovery gives a population of only 0.017. The tenth
order recovery gives a population of 0.90, which is nearly the noise-free
result. Defining an effective $T_1(n)$ for a
given order $n$ as the time the recovered evolution has
$1/e$ population remaining, we see $T_1(n)$ 
$\approx$ $(n+1) T_1(n=0)$. 

We also perform the relaxation time experiment on Rigetti's eight qubit chip,
Agave~\cite{reagor2018demonstration,smith2016practical}, a superconducting  qubit
quantum computer with a ring topology. Each of
the eight qubits has slightly different $T_1$ and $T_2^*$ times, all approximately 10
$\mu$s. Furthermore, these noise
characteristics drift in time~\cite{muller2015interacting,kandala2018extending}.
These features  provide the necessary variation in
noise parameters for our method. 
We first excite a single qubit using
a Pauli-X gate, wait some time, and measure the qubit state. 
This is repeated for many
different wait times. 
Each experiment at a given wait time is averaged over 10$^5$
shots, giving an average population. The
$T_1$ time and associated decay rate, $\gamma_1$, is
extracted by fitting an exponential to 
the data. 
This process is repeated for each qubit in the eight qubit chip, a few
minutes are allowed to pass, and then the full cycle is repeated, starting from the
first qubit. 
This generates many different repetitions with varying noise
parameters. The fitted $\gamma_1$ parameters are used for the
recovery method. Just as in the simulation, each wait time is recovered
separately. A total of 45 different single qubit repetitions
are run with results shown in Fig.~\ref{t1_experiment}.
Both first and second order recoveries 
result in a higher population than the average, similar to the
simulation results. The recovery is limited by noise sources  not included in
the model, such as readout noise. Neither first nor second order recovery
reaches the correct value of 1; instead, they recover $\approx$ 0.9,
roughly the average readout fidelity of the qubits used in the
experiment. 

\begin{figure}
  \centering
    \sidesubfloat[]{\includegraphics[width=\columnwidth]{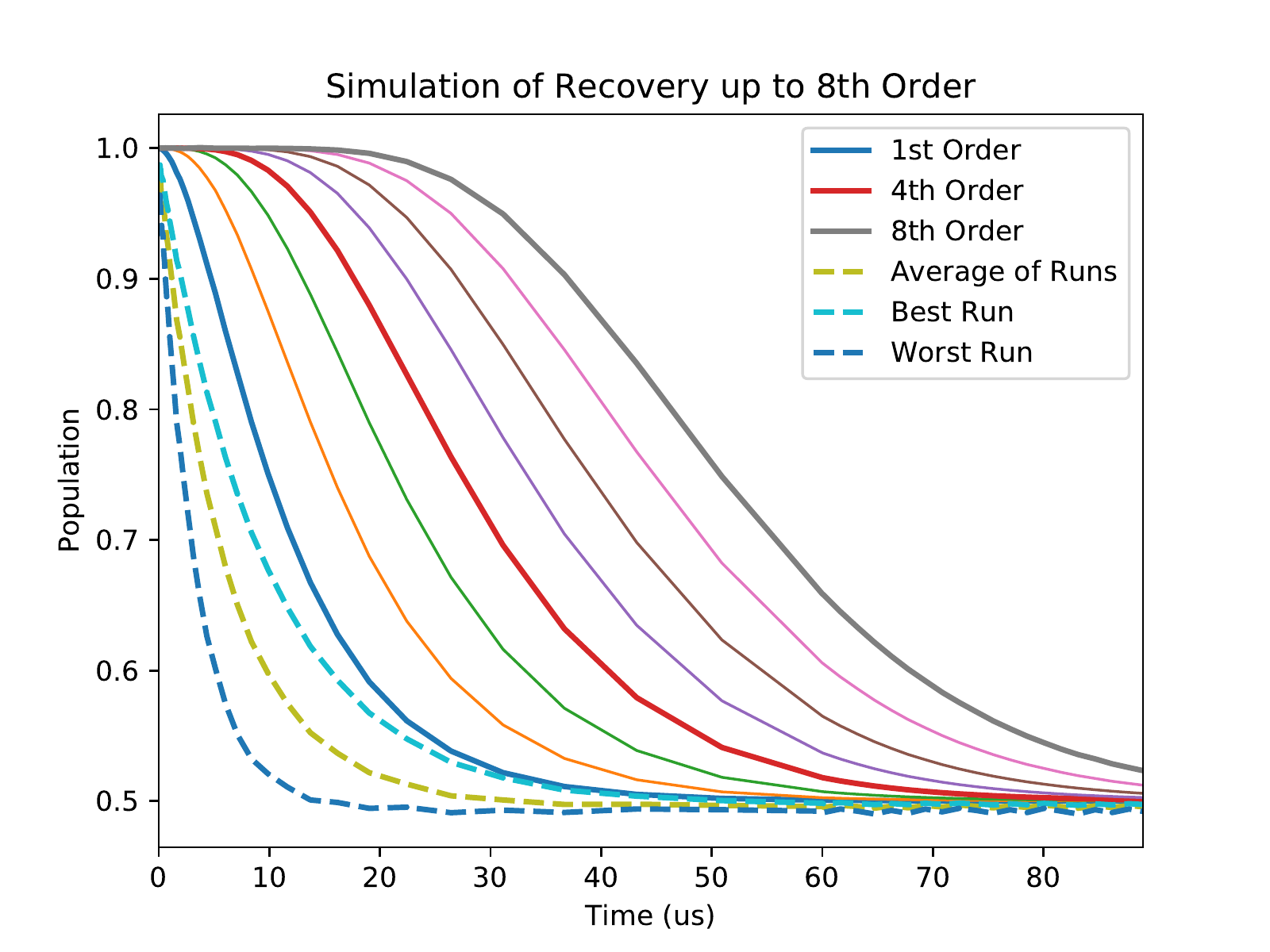}\label{t2_sim_8th}}
  \hfil
  \sidesubfloat[]{\includegraphics[width=\columnwidth]{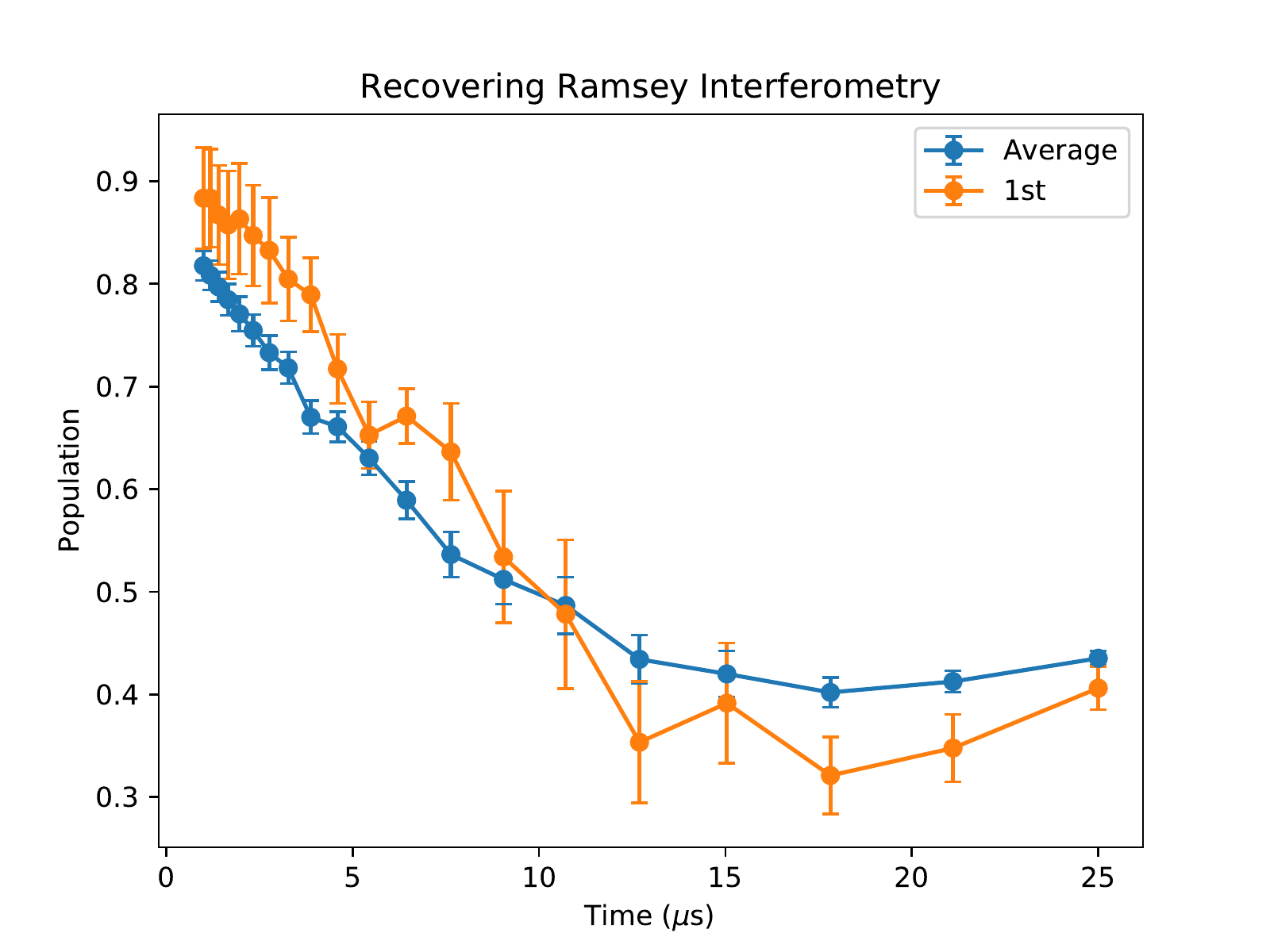}\label{t2_experiment}}
  \caption{Recovery of population in Ramsey interferometry experiment with no
    background magnetic field. (a) Simulated data from 450 simulations
       with random $T_1$ and $T_2^*$ times. (b) Experimental results on Rigetti's
      8 qubit quantum computer, 
      Agave~\cite{reagor2018demonstration}; 12 separate repetitions are included at
      each time, with experimentally determined $T_1$ and $T_2^*$ times. } \label{t2_figs}
\end{figure}

With only one noise parameter the method is
very similar to Richardson extrapolation, which has been shown to be able to
extrapolate to the zero-noise limit in 
superconducting
qubits~\cite{kandala2018extending,temme2017error};
mathematically, the methods differ only in the choice of points and specific
fitting strategy, but the source of the variation in noise differs
greatly. 
The Richardson extrapolation technique
assumes a single global noise source, implemented by scaling the length
of the pulses while running the quantum algorithm on
the same set of qubits~\cite{kandala2018extending}. 
The hypersurface method allows for any number of 
noise sources and can be thought of as a multi-dimensional
generalization which utilizes the natural variations in qubit properties. To
show this, we use Ramsey interferometry with  
no background magnetic field, a common technique for measuring $T_2$
times~\cite{paik2011observation}. This involves  
applying a $\pi/2$ $X$ rotation to $| 0 \rangle$, waiting some time, and then applying another
$\pi/2$ $X$
rotation. Without noise and in a rotating frame, the final state would be $| 1 \rangle$. As
opposed to the relaxation time experiment, 
both the amplitude damping and dephasing
noise affect the result. 
Figure~\ref{t2_sim_8th} shows simulated results of this experiment with
recovery up to 8th order. Spontaneous emission $T_1$ and pure dephasing 
$T_2^*$ times are independently, randomly
chosen between 5 $\mu s$ 
and 15 $\mu s$ with 450 simulations being run. 
Without recovery, the excited state population associated with a 
given $\gamma_1 = 1/T_1$ and $\gamma_2 = 1/T_2^*$  
exponentially decays in time with rate $1/T_2 = \gamma_1/2 + \gamma_2$.
As with
the relaxation time experiments, first order recovery leads to 
a better evolution for most points compared with even the best run of all the
qubits. As ever higher orders are considered, 
a unity excited state population is
recovered for longer periods of time.

We also carry out  Ramsey experiments on  Rigetti's eight qubit Agave
quantum computer. To obtain the correct noise rates, $\gamma_i$, for a given
repetition we first characterize $T_1$ using the relaxation time experiment
discussed above. 
We then perform a Ramsey interferometry experiment as previously described.
The results of this experiment are fit
to an exponential to obtain $T_2$.
This allows 
determination of the pure dephasing rate via 
$\gamma_2 = 1/T_2 - 1/(2T_1)$.
Each wait time in
both the $T_1$ and $T_2$ determinations is averaged over 10$^5$ shots. With
both necessary rates determined, the hypersurface method,
equation~\eqref{hs_lsq}, is used to recover the excited state population 
at each
wait time, using 12 
repetitions (Fig.~\ref{t2_experiment}). 
The first order correction recovers a significant fraction
of the missing population and approaches the readout
fidelity limit at early times.

Characterization of the noise rates and a good understanding of the
important noise sources is imperative for this method. For example,
because of the way $\gamma_2$ is
determined, poor determinations of $T_1$ and $T_2$ can sometimes lead to
negative $\gamma_2$; these unphysical repetitions are excluded from
the recovery in our experiments. 
In both experiments on the Rigetti quantum computer, the correct 
unity excited state population could not be 
recovered. Instead, the method recovered the
readout fidelity limit. In superconducting qubit systems, 
where noise
fluctuates over time scales in the few minute range, $T_1$ and $T_2$ would
need to be characterized before essentially every repetition to ensure that the
hypersurface equations have the correct noise sources.

The Ramsey experiments show the strength of this method, compared to Richardson
extrapolation with a global noise parameter. Rather than stretching out the
circuit, limiting the number of logical gates that can be done, our method
involves determination of the noise parameters (in this case, $T_1$ and $T_2^*$
times), followed by evaluation of the relevant algorithm or experiment. 
Furthermore, this
can be done in parallel. In superconducting circuits, each qubit
has slightly different noise characteristics. 
Our method allows these different noise characteristics to be utilized for the
recovery of the observable. For example, several small quantum computers could
be produced and an 
algorithm  could be evaluated in
parallel.
At the end, the results could be combined to give approximate noise-free
observables. The overhead of the method can be paid in
either time or space. 
If the quantum computer is much larger than the number of
logical qubits needed for a given algorithm, the extra qubits can be used to run
the same algorithm in parallel.
Alternatively, the
algorithm can be run many times in succession; here, the noise characteristics
can be changed by waiting for natural fluctuations, by changing the logical
to physical qubit mapping,
or by carefully tuning the noise (if
the quantum hardware allows that). By utilizing a single global
  noise source, the Richardson extrapolation technique generally requires many fewer
  evaluations than the hypersurface method; however, these evaluations must be
  done before the natural fluctuations of the underlying noise parameters
  change. Furthermore, in quantum sensors and general quantum experiments, it
  can be infeasible to stretch out the evolution; the hypersurface method only
  requires the same evolution to be repeated.

\begin{figure}
  \centering
  \includegraphics[width=\columnwidth]{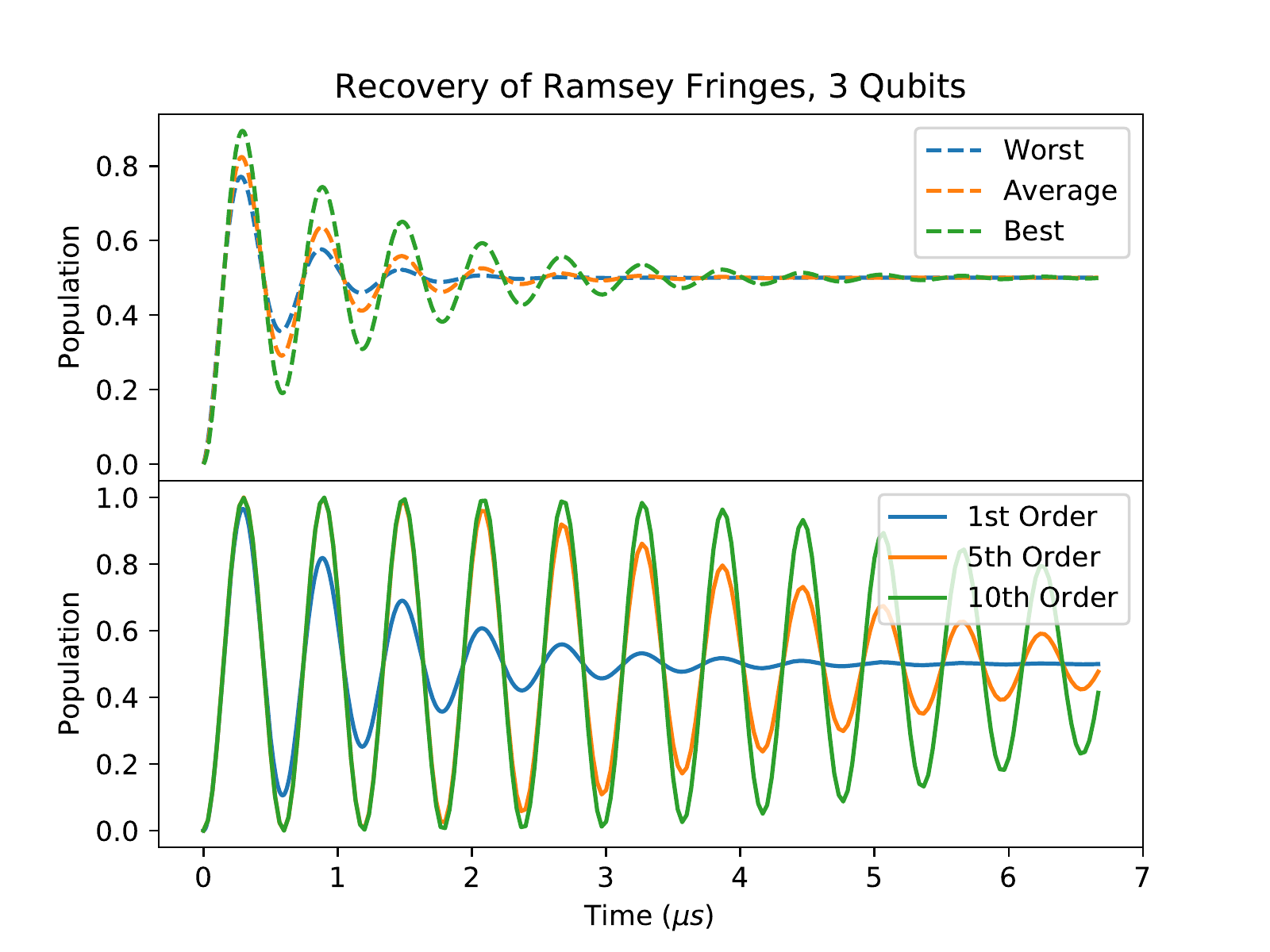}
  \caption{Recovery of Ramsey fringes from entangled qubits using 350 simulations of three qubits
    with random $T_2^*$ times.}\label{3q_ramsey}
\end{figure}

As a final example, we perform a simulation of Ramsey interferometry with a
background magnetic field. This is a common technique used in quantum
sensing of magnetic fields~\cite{degen2017quantum}. 
Both single qubit and entangled
arrays of qubits can be used as sensors~\cite{huelga1997improvement}. 
The system is first prepared in a superposition state such as
$\frac{1}{\sqrt{2}}(|0\rangle + |1\rangle)$ in the case of a single qubit sensor and the
GHZ state, $\frac{1}{\sqrt{2}}(|00\cdots 0\rangle + |11\cdots 1\rangle)$, for
an 
entangled array of qubits~\cite{degen2017quantum}. 
Single qubit superposition states can be prepared
with a Hadamard gate; GHZ states can be prepared 
with a Hadamard gate and a
sequence of CNOT gates.
The system  then evolves in the presence
of a background magnetic field, causing it to pick up a phase. 
The inverse 
entangling operation 
is then applied, transferring the phase onto a single
qubit, which is then measured. The phase accumulated over the course of the
interaction depends on the strength of the magnetic field and the time the
system interacts with it. 
If the interaction
time is too much longer than the coherence time, all of the useful information
decays away, and the characteristic Ramsey fringes will no longer be visible. 
We
simulate this experiment using both a single qubit and three 
entangled qubits. For each qubit, we select a random $T_2^*$ times in the range
of $0.5 \mu$s to $1.5 \mu$s, consistent with parameters for nitrogen-vacancy
centers~\cite{jelezko2004observation}. The characteristic $T_1$ time is large
enough to be safely ignored. We set the background 
magnetic field to 10 $\mu$T and use a total of 350 samples for both the single
qubit and entangled qubit simulations. Figure~\ref{3q_ramsey} shows the recovery
of Ramsey fringes from a three qubit GHZ state; the single qubit case is
plotted in the Appendix. In the three qubit case
we now have three noise parameters, one for each qubit. The higher order
recovery now involves many cross terms, and the resulting hypersurface is
three-dimensional. 
Even in this maximally entangled state, the Ramsey fringes are still 
recovered long after most of the individual fringes have decayed away.

We presented a novel method to recover arbitrary quantum
observables by repeatedly measuring the observables with differing
noise rates and fitting a hypersurface to the repetitions.
By including more and more repetitions and increasing the hypersurface order, an
increasingly good approximation of noise-free observables of general 
quantum system can be recovered. 
For many quantum systems, $T_1$ and $T_2^*$
noise are the dominant 
noise sources and there are many techniques for characterizing them. Our 
method can recover a good approximation of the noise-free evolution
in these cases. 
As shown in this Letter, this method has applications in quantum
computing and quantum sensing. 
Further study on ways to minimize the number of
repetitions needed for higher-order recovery of a large number of coupled
quantum systems is necessary to control the overhead of the method.
Demonstrations of more complicated algorithms and larger numbers of coupled
quantum systems is
also warranted. 

\begin{acknowledgments}
This work was performed at the Center for Nanoscale Materials, a U.S. Department
of Energy Office of Science User Facility, and supported by the U.S. Department
of Energy, Office of Science, under Contract No. DE-AC02-06CH11357. We
gratefully acknowledge the computing resources provided on Bebop,
a high-performance computing cluster operated by the Laboratory Computing
Resource Center at Argonne National Laboratory. We thank Rigetti Computing for
use of the Agave quantum computer, and Guenevere Prawiroatmodjo for help in
setting up the experiments on Agave.
\end{acknowledgments}

\clearpage
\appendix
\section{Appendix}
\subsection{Recovery of Single Qubit Ramsey Fringes}
\begin{figure}
  \centering
  \includegraphics[width=\columnwidth]{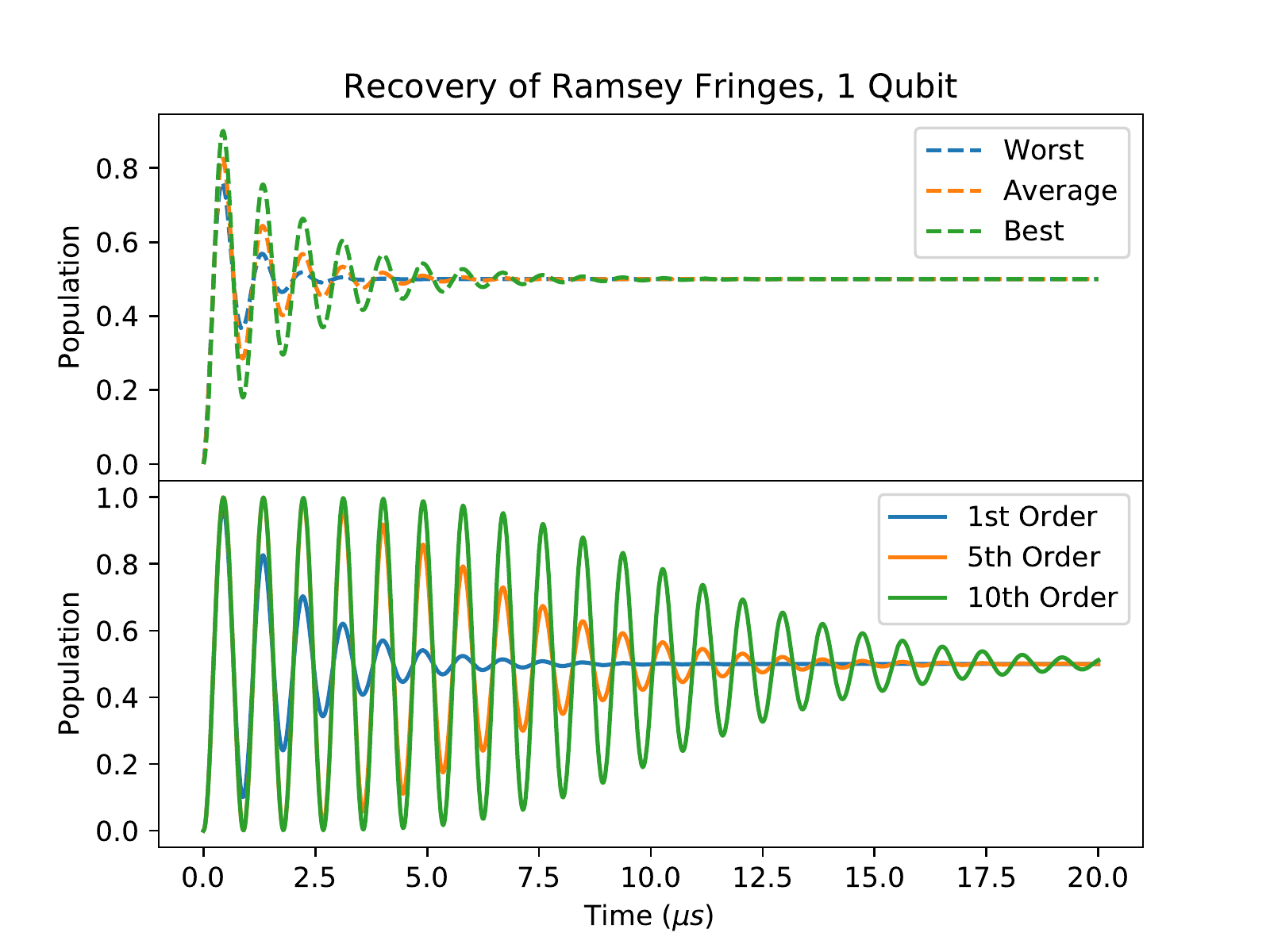}
  \caption{Single qubit Ramsey interferometry. 350 simulations with random
    $T_2^*$ times are used. }\label{1q_ramsey}
\end{figure}

 Figure~\ref{1q_ramsey} shows the recovery
 of Ramsey fringes for a single qubit. In this single qubit case, there is only
 one noise parameter, $\gamma_2$, as $\gamma_1\ll \gamma_2$, in contrast
 to the Ramsey interferometry in the superconducting qubit system, where
 $\gamma_1 \approx \gamma_2$. We set the background 
 magnetic field to 10 $\mu$T and combine the results of 350 experiments. As the
 order of the recovery is  increased, more and more fringes are recovered; at
 tenth order, the fringes 
 extend are recovered even when the best single qubit run has no clearly visible
 fringes. Without correction, the fringes decay with the characteristic $T_2$
 time~\cite{paik2011observation}.
 
\subsection{Derivation}
The Lindblad master equation for a general density matrix $\rho(t)$ evolved from
an initial state $\rho_0$ is defined as (throughout, $\hbar$ = 1)
\begin{equation}\label{lindblad_me}
  \frac{d\rho}{dt} = -i [H, \rho_0] + L(\rho_0),
\end{equation}
where $H$ is the Hamiltonian of the system, describing coherent evolution and
$L$ is the Lindblad superoperator, describing incoherent evolution, such as
noise processes. Both $H$ and $L$ can, generally, be time dependent.
`Vectorizing' the density matrix~\cite{uzdin2015}, $\rho$, allows us to write a general
solution of the Lindblad master equation as
\begin{equation}\label{lme_sol}
  \tilde\rho(t) = \exp(-i\tilde H t + \tilde L t) \tilde\rho_0,
\end{equation}
where $\tilde\rho$ is the vectorized density matrix, $\tilde H$ is the vectorized
Hamiltonian, and $\tilde L$ is the vectorized Lindblad superoperator. See Ref. \onlinecite{uzdin2015} for
more details about the process of vectorization. We can decompose the exponential in
the solution, equation~\eqref{lme_sol}, using the Trotter decomposition~\cite{trotter1959product}:
\begin{equation}\label{trotter}
  \exp(-i\tilde H t + \tilde L t) = \lim_{n\rightarrow \infty}\Bigg(\Big(\exp(-i\tilde H \frac{t}{n}) \exp(\tilde L \frac{t}{n})\Big)^n\Bigg)
\end{equation}
and take the Taylor expansion of the exponential of the Lindblad
\begin{equation}\label{trotter_taylor}
  \exp(-i\tilde H t + \tilde L t) = \lim_{n\rightarrow \infty}\Bigg(\Big(\exp(-i\tilde H \frac{t}{n}) \sum_{m=0}^\infty \frac{(\tilde L \frac{t}{n})^m}{m!}\Big)^n\Bigg).
\end{equation}
We now write the Lindblad superoperator as a sum of many different Lindblad
superoperators, each with its own rate: $\tilde L = \sum_j \gamma_j \tilde L_j$ and plug
this into equation~\eqref{trotter_taylor}, giving
\begin{equation}\label{tt_l_expanded}
  \exp(-i\tilde H t + \tilde L t) = \lim_{n\rightarrow \infty}\Bigg(\Big(\exp(-i\tilde H \frac{t}{n}) \sum_{m=0}^\infty \frac{(\sum_j \gamma_j \tilde L_j \frac{t}{n})^m}{m!}\Big)^n\Bigg).
\end{equation}
Equation~\eqref{tt_l_expanded} an infinite
sum over $m$, a finite sum over $i$, and is raised to an infinite power, $n$.
Though this equation has infinite 
terms, we can collect all terms that have the same $\gamma$
prefactors. For example, the collection of all terms with only $\gamma_j$ will
include all terms from the product that have one first order element from
the Taylor series expansion of $\tilde L$. Terms with $\gamma_j^2$ will include
products with two first order elements, as well as products with one second
order element. To provide a concrete
example, we truncate the Trotterization at third order, the Taylor expansion at
fist order, and include two noise terms. Let $U=\exp(-i \tilde H \frac{t}{3})$
and $V_j = \tilde L_j \frac{t}{3}$. With our truncation, we rewrite
equation~\eqref{tt_l_expanded} as
\begin{widetext}
\begin{equation}\label{3rd_order_Trotter}
  \begin{split}
    \exp(-i\tilde H t + \tilde L t) &\approx \Big(U (1 + \gamma_1 V_1 + \gamma_2 V_2)\Big)^3 \\
    &\approx U U U + \gamma_1 (U V_1 U U + U U V_1 U + U U U V_1)
    + \gamma_2 (U V_2 U U + U U V_2 U + U U U V_2) + \\
    &~~+\gamma_1^2 (U V_1 U V_1 U + U U V_1 U V_1 + U V_1 U U V_1)
    + \gamma_2^2 (U V_2 U V_2 U + U U V_2 U V_2 + U V_2 U U V_2) \\
    &~~+ \gamma_1 \gamma_2 (U V_1 U V_2 U + U U V_1 U V_2 + U V_1 U U V_2
    + U V_2 U V_1 U + U U V_2 U V_1 + U V_2 U U V_1).
  \end{split}
\end{equation}
\end{widetext}
The generalization to higher order Trotterizations is clear; the expanded sum
will have many more terms (due to each term having a smaller timestep,
$\frac{t}{n}$), but terms can be grouped by their $\gamma_j$ prefactors. For
higher order Taylor expansions of $\exp{\tilde L}$, terms can still be grouped
by their $\gamma_j$ prefactors. For example, take the $\gamma_1^2$ term from
equation~\eqref{3rd_order_Trotter}. With a second order Taylor expansion, the
$\gamma_1^2$ terms now contain contributions from $V_i^2$:
\begin{equation}
  \begin{split}
    [\gamma_1^2~\text{terms}] &= U V_1 U V_1 U + U U V_1 U V_1 \\
    &~~+ U V_1 U U V_1 + U V_1^2  U U \\
    &~~+ U U V_1^2 U + U U U V_1^2
  \end{split}
\end{equation}
Combinging the generalizations to both higher order Trotterization and higher
order Taylor expansion is relatively straightforward; the number of terms grows
precipitously, but they can always be gathered by their $\gamma_j$ prefactors.
Collecting all the terms for both the infinite limits of Trotterization and the
Taylor expansion leads to
\begin{equation}\label{collected}
  \begin{split}
    \exp(-i\tilde H t + \tilde L t) &= \lim_{n\rightarrow \infty}\Bigg(\Big(\exp(-i\tilde H \frac{t}{n})\Big)^n\Bigg) \\
    &+ \sum_j \gamma_j [\gamma_j~\text{terms}] \\
    &+ \sum_j \sum_k \gamma_j \gamma_k[\gamma_j\gamma_k~\text{terms}] + \cdots,
  \end{split}
\end{equation}
where we have used $[\gamma_j~\text{terms}]$ to represent the (infinite)
collection of terms all with $\gamma_j$ as a prefactor.
This represents the general (exact) evolution operator for our system. 
Our density matrix at
time $t$ can now be written
\begin{equation}\label{final_evolution}
  \begin{split}
    \tilde\rho(t) &= \lim_{n\rightarrow \infty}\Bigg(\Big(\exp(-i\tilde H \frac{t}{n})\Big)^n\Bigg) \tilde\rho_0\\
    &+\sum_j \gamma_j [\gamma_j~\text{terms}] \tilde\rho_0\\
    &+\sum_j \sum_k \gamma_j \gamma_k[\gamma_j \gamma_k~\text{terms}] \tilde\rho_0 + \cdots.
  \end{split}
\end{equation}
The first term of this expansion represents the noise-free result. Other
terms represent the effects of noise on the evolution. Since this method
directly corrects the density matrix, $\rho$, it follows that it also corrects
an arbitrary observable,
\begin{equation}\label{final_evolution}
  \langle A \rangle = A_0 +\sum_j \gamma_j A_{j} + \sum_j \sum_k \gamma_j \gamma_k A_{jk} + \cdots,
\end{equation}
where $A_0$ is the noise-free observable value and $A_{j}$ is the effect of
noise rate $j$ on the observable.
We define the `order' of
the combined Trotterization and Taylor expansion by the number of $\gamma_j$
terms included. The first order terms, for example, are those with a single
$\gamma_j$ prefactor, and include all possible ways that one (infinitesimal)
error evolution can be included. The second order terms include all possible ways
that two (infinitesimal) error evolutions can be included, and so on. We do not
{\em a priori}
know what the values of the effects of noise on the observable (such as $A_j$)
for any order are;
however, we can characterize $\gamma_j$ 
for a given experiment.
By taking a sequence of experiments, varying $\gamma_j$, we
can reconstruct the unknown evolution terms by fitting a hypersurface to the
points. The coefficients of the hypersurface represent the effects (or, for the
zeroth order term, the lack of effects) of noise to a given order on
the density matrix (or an arbitrary observable). 
Equation~\eqref{final_evolution} generally has effects up the infinite order; to
make it tractable, we truncate at some given order. As more orders are included,
the fit becomes more accurate, and, therefore, a better approximation of the
noise-free result is obtained.

\subsection{Number of Terms in Expansion}
Naively, the number of terms in a given order $l$ would be $m^l$, where $m$
is the number of noise terms (which could be the number of qubits or a small
factor times the number of qubits). Since $\gamma_j$ is a scalar, all $\gamma_j$
will commute, allowing us to fuse terms with the same set of $\gamma$. For
instance, given a second order expansion with two noise terms, we would general
have terms with prefactors $\gamma_1 \gamma_1, \gamma_1 \gamma_2, \gamma_2
\gamma_1,$ and $\gamma_2 \gamma_2$, but since $\gamma_1 \gamma_2 = \gamma_2
\gamma_1$, we can reduce the number of fitted parameters by combining the
$\tilde L_1 \tilde L_2$ and $\tilde L_2 \tilde L_1$ terms. For a given order $l$
and number of noise terms $m$, the number of parameters $n$ for that order is
the number of
multinomial coefficients, which is given by the formula
\begin{equation}\label{multinomial}
  n = \binom{l+m-1}{m-1}.
\end{equation}
For an expansion truncated at order $l$, the total number of parameters, for all
orders, is the sum of equation~\eqref{multinomial} for each order up to, and
including, $l$.

\bibliography{hs_hqec}
%
\end{document}